\documentclass[11pt]{article}

\usepackage{psfig}
\usepackage{epsfig}
\usepackage{color}

\def\be{\begin{equation}}
\def\ee{\end{equation}}
\def\bea{\begin{eqnarray}}
\def\eea{\end{eqnarray}}

\begin{document}

\vspace*{-3cm}
{\flushright{\bf LAL 99-54}\\
\hspace*{12.5cm}
{October 1999}}
\vskip 1 cm

\vspace*{2.5cm}
\centerline {\LARGE\bf New Structure Function Results}
\vspace{2mm}
\centerline {\LARGE\bf at Low $x$ and High $Q^2$ from HERA\footnote{Talk given
at Hadron'99, Aug.\ 24-28, 1999, Beijing, China}}
\vspace{1.5cm}
\centerline {\bf\Large Zhiqing ZHANG}

\vspace{1cm}
\centerline {\bf\large Laboratoire de l'Acc\'el\'erateur Lin\'eaire,}
\centerline {\it IN2P3-CNRS et Universit\'e de Paris-Sud, BP 34,
F-91898 Orsay Cedex}
\centerline {\it E-mail: zhangzq@lal.in2p3.fr}   

\vspace{3cm}

\begin{abstract}
A precise proton structure function $F_2$ at low $x$ is measured.
The data is interpreted in the framework of QCD with an extraction of gluon
density $xg$. 
The charm contribution to $F_2$ is determined in an extended kinematic range.
Neutral and charged current cross-sections at high
$Q^2$ are also measured and compared with the Standard Model predictions.
\end{abstract}


\section{INTRODUCTION}

The deep inelastic scattering (DIS) experiments have played an important
role in establishing and developing the quantum chromodynamics
(QCD) and electroweak (EW) sector of the Standard Model (SM).
At HERA, the first $e^\pm p$ collider operating at a centre-of-mass energy of 
$\sqrt{s}=300-320$\,GeV, the neutral current (NC)
and charged current (CC) DIS processes are being studied since 1992
with increasing precision by the H1 and ZEUS experiments in a largely new
kinematic domain at low $x$ and high $Q^2$.
The experimentally measurable kinematic variable $x$ stands for the momentum
fraction carried by the struck parton in the quark-parton model, and $Q^2$
the negative of the four-momentum transfer squared of the exchanged bosons.

\section{SELECTED RECENT MEASUREMENTS AT LOW $x$ AND HIGH $Q^2$}

In this talk, a few recent measurements at low $x$ and high $Q^2$ are
reported\footnote{Due to the space limitation, a further
selection has been made from what have actually been shown during the
conference.}: precision proton structure function $F_2$, gluon density
$xg$ from a next-to-leading-order (NLO) QCD analysis of $F_2$, 
charm contribution to $F_2$,
longitudinal structure function $F_L$, and NC
and CC cross-sections at high $Q^2$.

What is measured experimentally is the cross-section. For the NC process,
the cross-section can be expressed in terms of $F_2$, $F_L$, and parity
violating term $xF_3$:
\begin{equation}
\frac{d^2\sigma^\pm_{\rm
NC}}{dxdQ^2}=\frac{2\pi\alpha^2}{xQ^2}\left[Y_+F_2(x,Q^2)-y^2F_L(x,Q^2)\mp 
Y_-xF_3(x,Q^2)\right]
\end{equation}
where $Y_\pm =1\pm (1-y)^2$ is the helicity function with $y=Q^2/sx$ and
$F_2=F^{\rm em}_2(1+\delta_Z)$.
The dominant contribution $F^{\rm em}_2$ arises from the
photon exchange and the term $\delta_Z$ accounts for contributions from 
the $\gamma Z$ interference and the $Z$ exchange.
At $Q^2$ well below $M^2_Z$, both $\delta_Z$ and $xF_3$ are negligible. 
The second term in Eq.(1) is also negligible 
if $y$ is not too large ($y\leq 0.6$). Similarly for the CC process, 
the cross-section can also be related to the corresponding structure functions.
In the leading-order approximation this is given by:
\begin{eqnarray}
\frac{d^2\sigma^+_{\rm CC}}{dxdQ^2}=\frac{G_F^2}{4\pi
x}\left(\frac{M^2_W}{M^2_W+Q^2}\right)^2\left[x(\overline{u}+\overline{c})+(1-y)^2x(d+s)\right]\,,&&\\
\frac{d^2\sigma^-_{\rm CC}}{dxdQ^2}=\frac{G_F^2}{4\pi
x}\left(\frac{M^2_W}{M^2_W+Q^2}\right)^2\left[x(u+c)+(1-y)^2x(\overline{d}+\overline{s})\right]\,.
\end{eqnarray}
where $G_F$ is the Fermi coupling constant. Contrary to the NC process, 
where $F^{\rm em}_2$ is independent of the lepton beam polarity, the $e^\pm p$
CC processes probe different quark types.

\begin{figure}
\begin{center}
\setlength{\unitlength}{1mm}
\begin{picture}(160,190)(0,-190)
\put(5,-205){
\epsfig{file=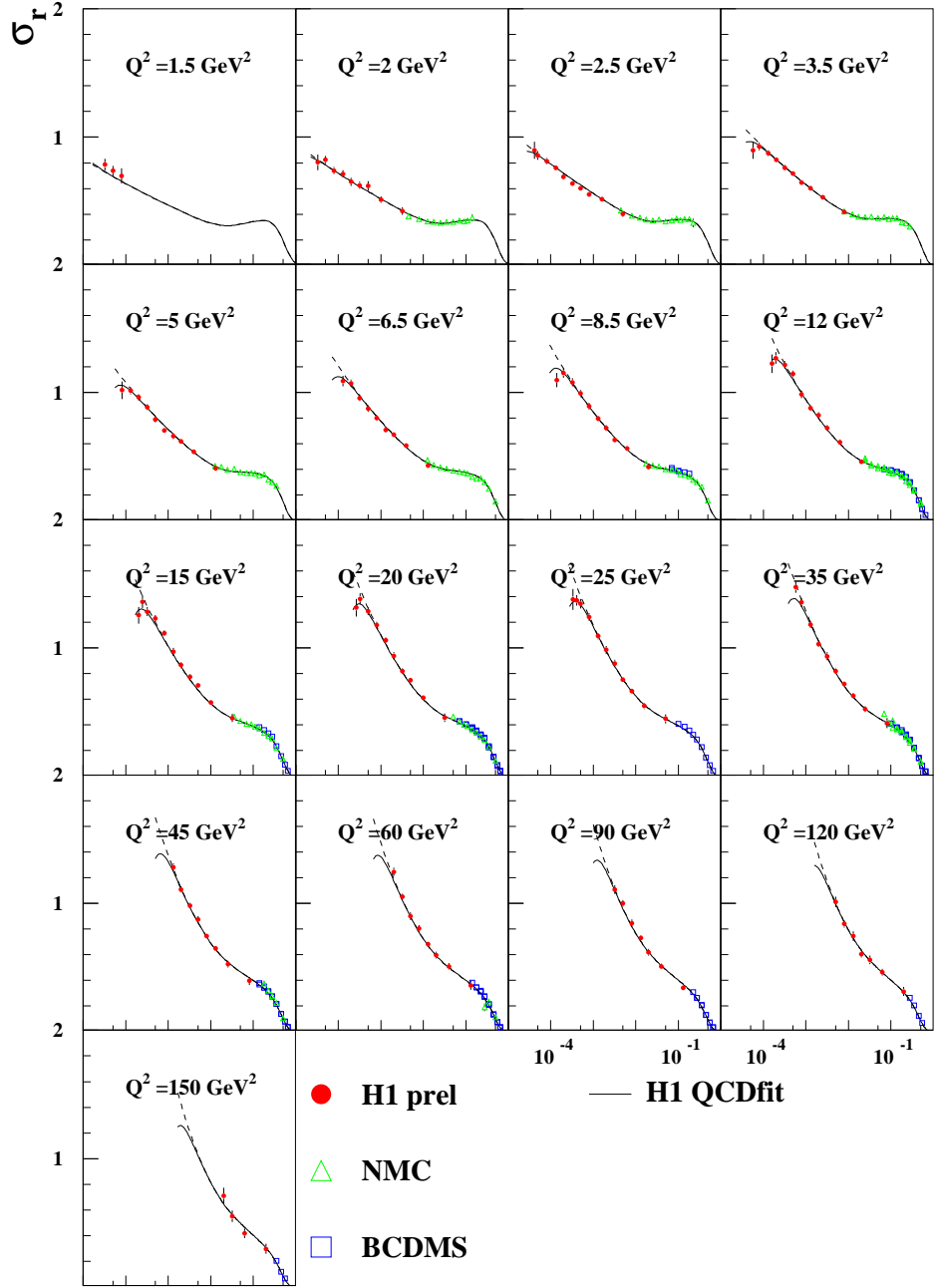,
bbllx=0pt,bblly=0pt,bburx=594pt,bbury=842pt,width=15cm}}
\end{picture}
\end{center}
\caption{H1 preliminary reduced cross-section compared with the H1 QCD fit 
(full curves) and with $F^{\rm em}_2$ (dashed curves).}
\label{fig:f2_x}
\end{figure}

The most precisely measured reduced cross-section\footnote{The
reduced cross-section is defined so that the $(x,Q^2)$ dependence on the
kinematic factor is removed, e.g.\ $\sigma_r=\tilde{\sigma}_{\rm
NC}=F_2(x,Q^2)-y^2/Y_+F_L(x,Q^2)\mp Y_-/Y_+xF_3(x,Q^2)$.} $\sigma_r$ sa far
at HERA is shown 
in Fig.\ref{fig:f2_x} as a function of $x$ for a fixed $Q^2$ value ranging from 1.5\,GeV$^2$ to 150\,GeV$^2$.

The H1 measurement is based on their
18\,pb$^{-1}$ $e^+p$ data taken in 1996 and 1997 for $Q^2>10\,{\rm GeV}^2$ and
a dedicated run of 2\,pb$^{-1}$ for $Q^2<10\,{\rm GeV}^2$~\cite{h1newf2}.
The statistical precision at $Q^2<100\,{\rm GeV}^2$ is now better than
1\% with a typical systematic precision of $3-4$\%.
The H1 data at $y\le 0.6$ together with fixed target data~\cite{nmc} at high
$x$ have been analysed using the NLO DGLAP evolution equations~\cite{dglap} to
provide predictions (H1 QCD fit) for $\sigma_r$ (full curves) and 
$F^{\rm em}_2$ (dashed curves).
The data at high $y$, which do not enter the fit, agree well with the 
predicted $\sigma_r$ but deviate from $F^{\rm em}_2$ showing the sensitivity 
to $F_L$. This sensitivity has been exploited by H1 with an extracted 
$F_L$ as shown in Fig.2.

\begin{figure}
\vspace*{-3.4cm}
\begin{center}
\setlength{\unitlength}{1mm}
\begin{picture}(160,80)(0,-80)
\put(-27.5,-185){
\hspace*{1.5cm}\epsfig{file=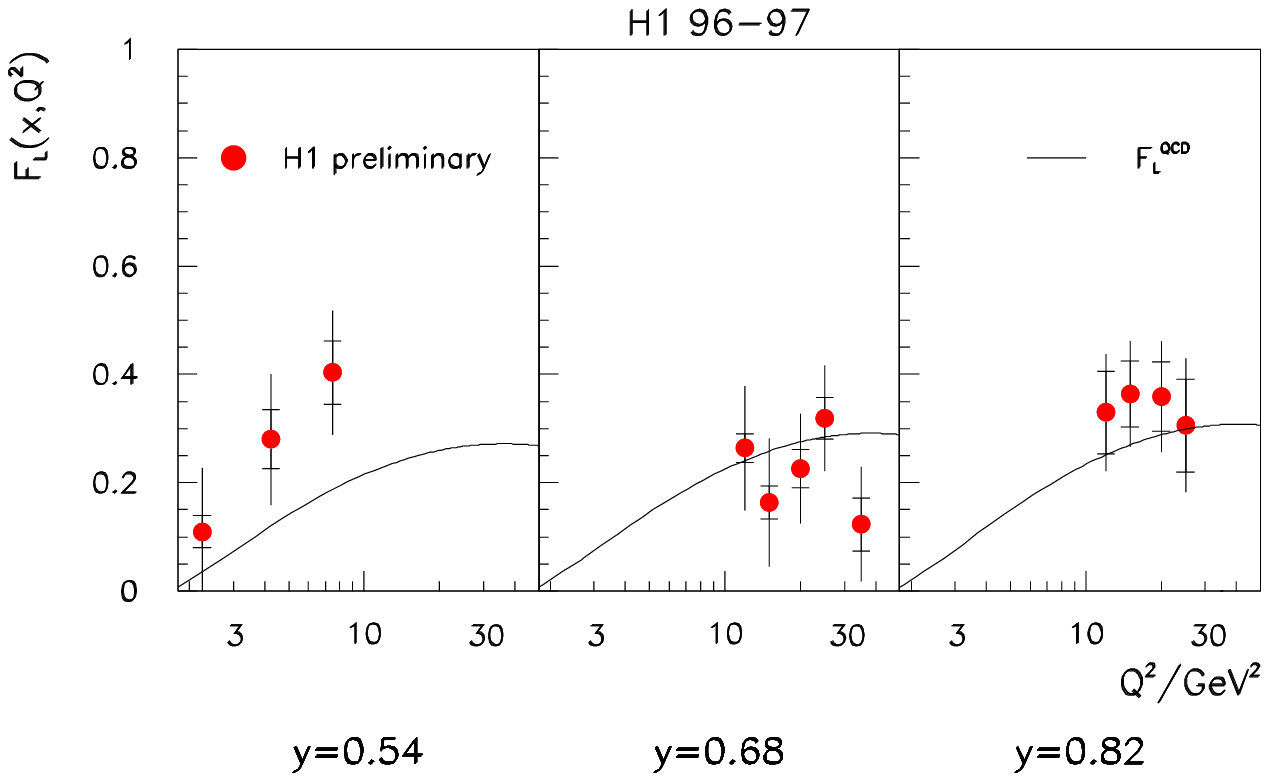,
bbllx=0pt,bblly=0pt,bburx=594pt,bbury=842pt,width=17cm}}
\end{picture}
\end{center}
\vspace*{1.5cm}
\caption{A determination of $F_L$ (H1 preliminary) compared with the QCD
prediction.}
\label{fig:fl}

\vspace*{-4.5cm}
\begin{center}
\setlength{\unitlength}{1mm}
\begin{picture}(160,150)(0,-150)
\put(-12.5,-200){
\hspace*{1.5cm}\epsfig{file=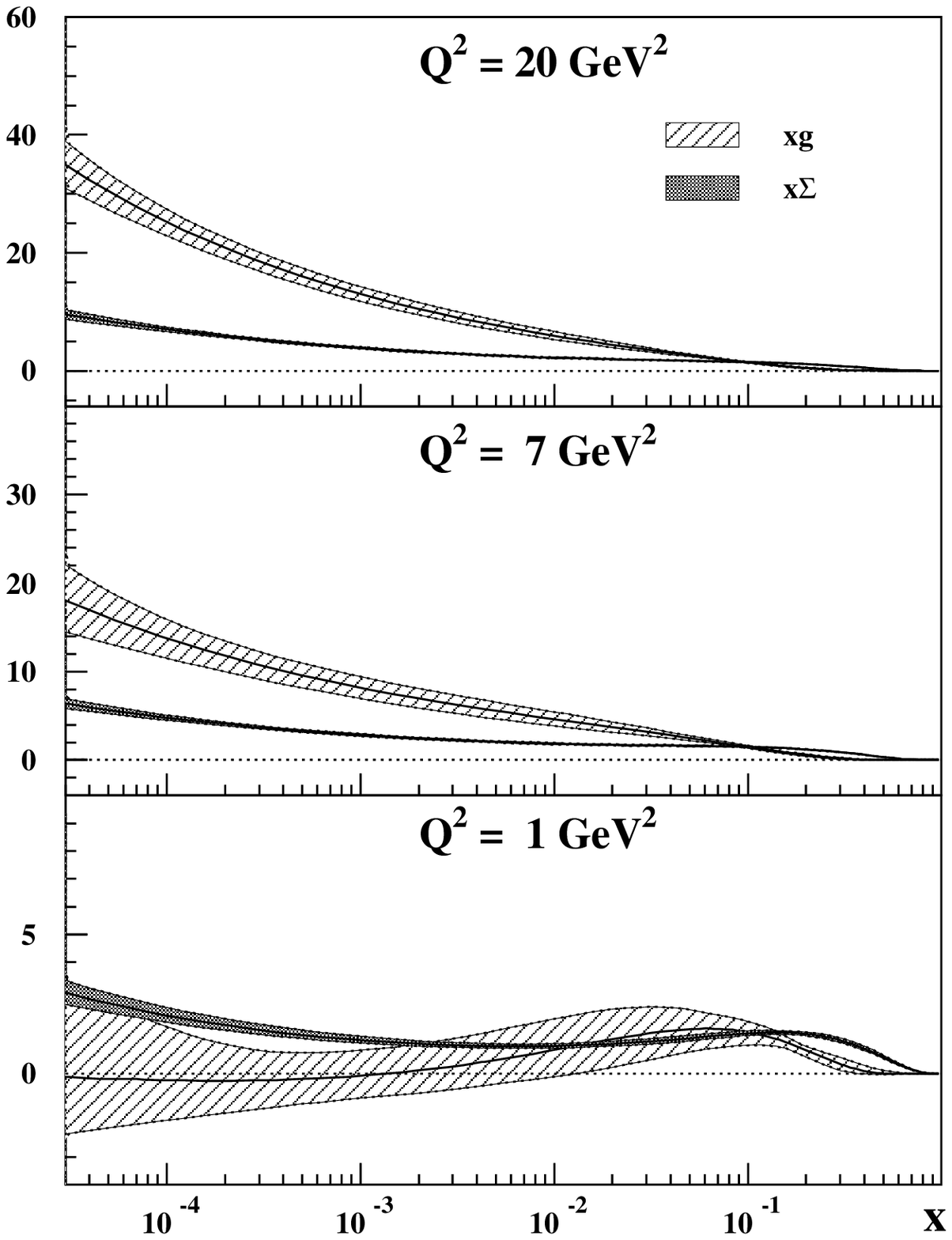,
bbllx=0pt,bblly=0pt,bburx=594pt,bbury=842pt,width=14cm}}
\end{picture}
\end{center}
\vspace*{0.5cm}
\caption{The results of the ZEUS QCD fit: the gluon momentum distribution $xg$
and the quark singlet momentum distribution $x\Sigma$ as a function of $x$.}
\label{fig:f2_gluon}
\end{figure}

The HERA data at low $x$ are unique as they can be used to extract the gluon
density $xg$.

The results of the ZEUS analysis are presented in Fig.3 
based on their previous $F_2$ data~\cite{f2_zeus_94}.
The relative magnitudes show that the rise of $F_2$ at low $x$ for $Q^2$
around 1\,GeV$^2$ is actually dominated by the quark singlet distribution
$x\Sigma\equiv \sum_{i=u,d,s}[xq_i(x)+x\overline{q}_i(x)]$ while the rise
for $Q^2$ at higher values can be attributed to the dominant gluon
distribution $xg$.

The charm production mechanism can be tested experimentally using tagged
$D^\ast$ decaying via $D^\ast\rightarrow D^0\pi\rightarrow (K\pi)\pi$. The
measured inclusive $D^\ast$ production cross-section~\cite{ds_xs}
can be described by NLO prediction~\cite{ds_pred} for charm
produced via photon-gluon fusion. Extrapolating the
measured cross-section into the full phase space, the open charm
contribution to $F_2$, $F^{c\overline{c}}_2$, has been determined. 
The results reported here are from ZEUS using their 1996-1997 data
(37\,pb$^{-1}$), a more than tenfold increase compared to the previous
studies~\cite{ds_94}. The ratio of $F^{c\overline{c}}_2$ over $F_2$ is shown in
Fig.4. The $F^{c\overline{c}}_2$, which contributes about 20\%
to $F_2$, increases as $Q^2$ grows and as $x$ falls, and agrees well with the
NLO prediction.

\begin{figure}
\begin{center}
\setlength{\unitlength}{1mm}
\begin{picture}(160,150)(0,-150)
\put(17.5,-160){
\epsfig{file=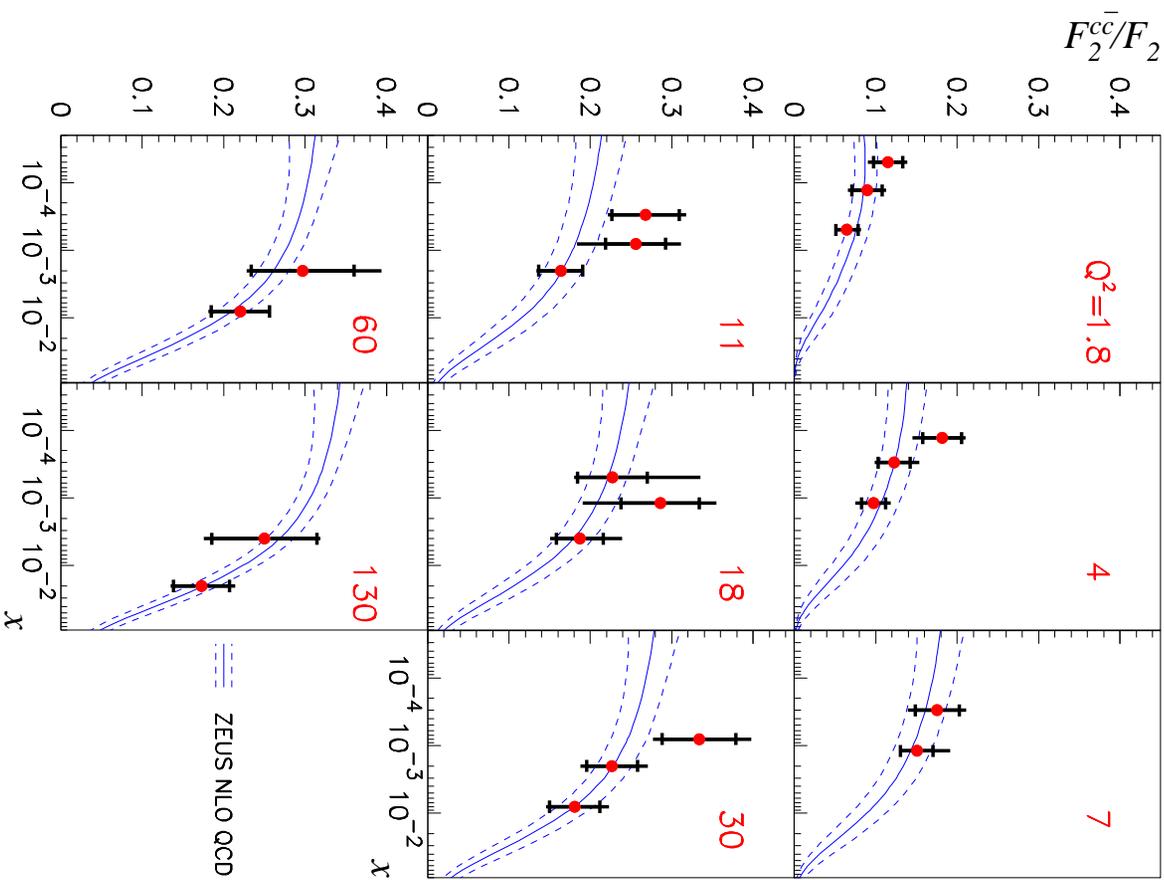,
bbllx=0pt,bblly=0pt,bburx=594pt,bbury=842pt,width=18cm}}
\end{picture}
\end{center}
\vspace*{1cm}
\caption{The ratio of the measured charm contribution $F^{c\overline{c}}_2$
over $F_2$ compared with the QCD prediction.}
\label{fig:f2cf2}
\end{figure}

The charm production data have further been used by H1 to 
determine in an alternative way the gluon density $xg$~\cite{ds_xs}. 
These results, though less precise, provide a powerful check of 
the $xg$ determined from the QCD analysis of the $F_2$ data.


The NC cross-sections at high $Q^2$ have been obtained by the H1 and ZEUS
collaborations using fully available $e^\pm p$ data~\cite{nc_hiq2_e+-p}.
The H1 results cover $150\,{\rm GeV}^2\leq Q^2\leq 30000\,{\rm GeV}^2$ and
$0.0032\leq x\leq 0.65$. The results at high $x$
are shown in Fig.5. The data are compared with predictions
derived from a slightly different fit~\cite{nc_hiq2_e+-p} from the one
mentioned earlier but using also only low $Q^2$ $e^+p$ data.
The data, though limited in statistical precision, agree fairly well with
the predictions except for a systematic excess of the $e^+p$ data at $x=0.4$
and $Q^2>15000\,{\rm GeV}^2$ and a step behaviour at $x=0.65$ between the H1
and the fixed target data. The excess, though less significant with
respect to what has been reported earlier based on 1994-1996 $e^+p$
data~\cite{excess_hiq2}, remains to be clarified.
The data also present a clear evidence for 
the EW effect associated with the $Z$ exchange
at high $Q^2$ in the NC process. 

\begin{figure}
\vspace*{-8cm}
\begin{center}
\setlength{\unitlength}{1mm}
\begin{picture}(160,165)(0,-165)
\put(-7.5,-167.5){
\hspace*{2.5cm}\epsfig{file=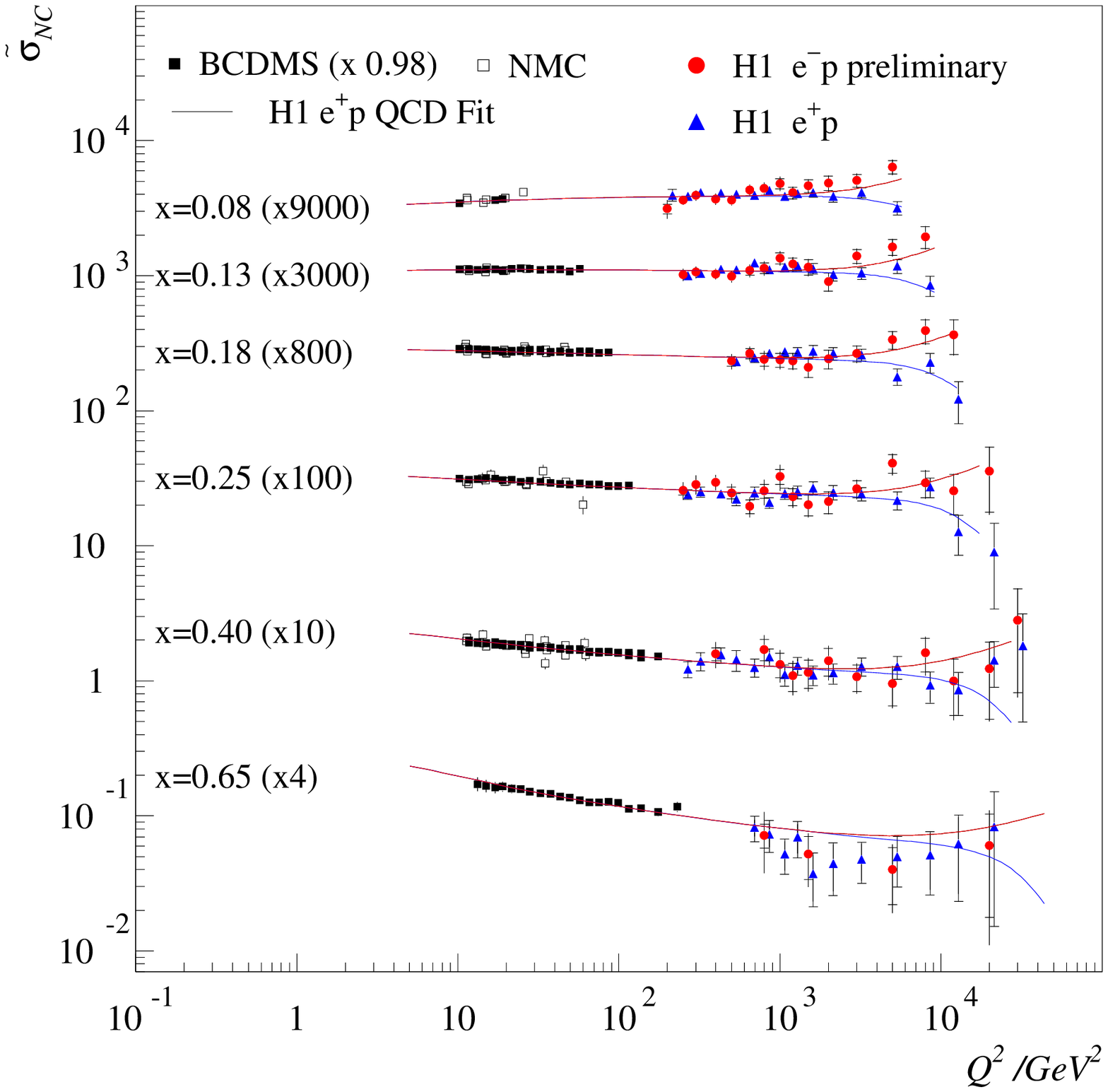, width=9.5cm}}
\end{picture}
\end{center}
\caption{H1 reduced $e^+p$ NC cross-section and preliminary reduced $e^-p$
NC cross-section compared with H1 $e^+p$ QCD fit.}
\label{fig:h1_hix}

\vspace*{-10cm}
\begin{center}
\setlength{\unitlength}{1mm}
\begin{picture}(160,165)(0,-165)
\put(-7.5,-210){
\hspace*{2.5cm}\epsfig{file=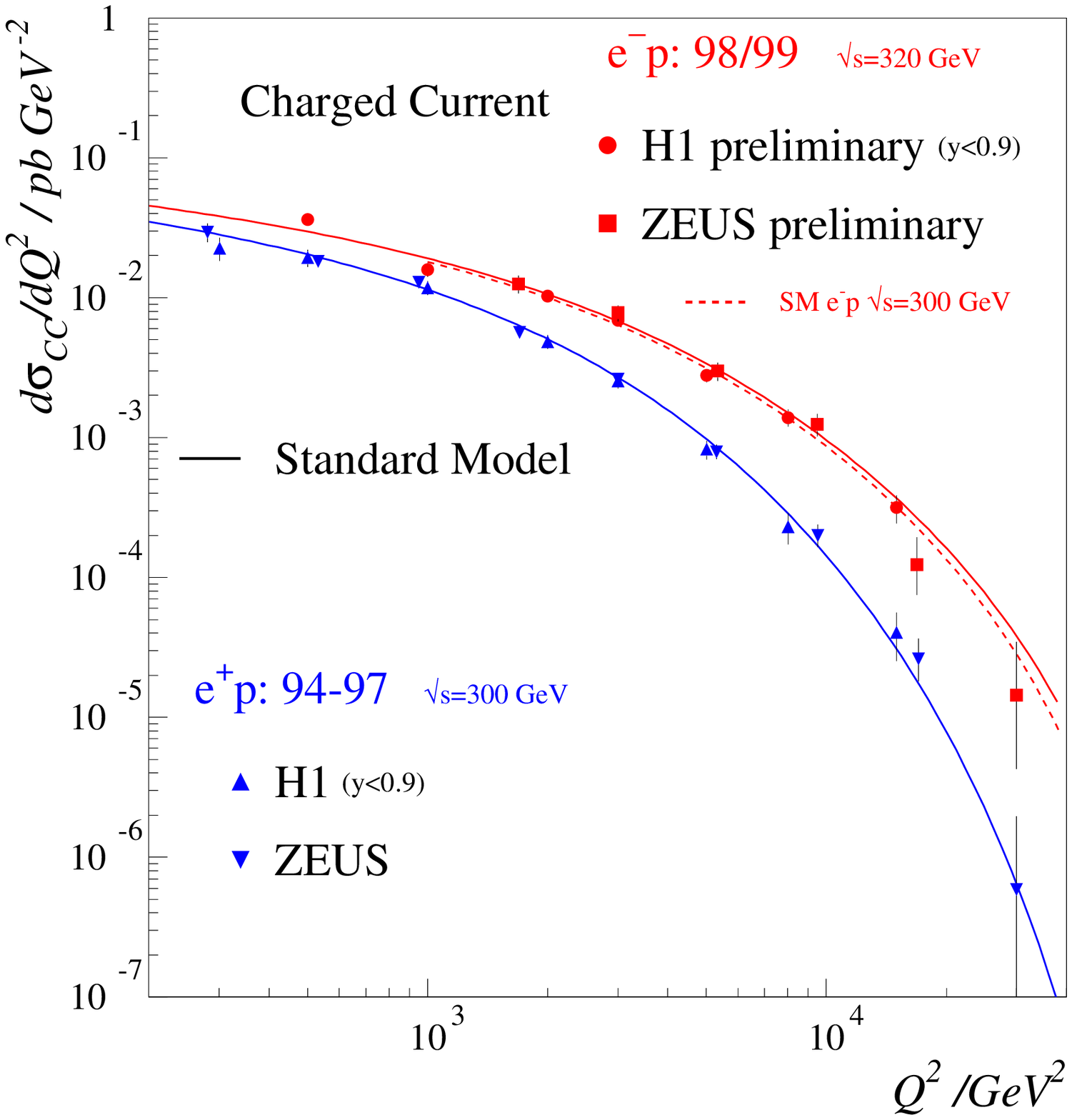,
bbllx=0pt,bblly=0pt,bburx=594pt,bbury=842pt,width=10cm}}
\end{picture}
\end{center}
\vspace*{1.5cm}
\caption{H1 and ZEUS $e^+p$ CC cross-sections and preliminary $e^-p$
CC cross-sections compared with SM predictions.}
\label{fig:cc_dsdq2}
\end{figure}

The CC cross-sections have also been measured~\cite{cc_hiq2_e+-p}.
The difference in the $e^+p$ and $e^-p$ scatterings shows the large
selectivity in the lepton quark coupling and the 
potential to constrain $u$ and $d$ flavours in a region that is free of
nuclear binding effects.


\section{SUMMARY AND OUTLOOK}

To summarise,
the structure function $F_2$ measured by the HERA experiments at low $x$ and 
$Q^2$ has now reached a precision comparable with that of fixed
target experiments. The gluon density determined from a NLO QCD analysis of
$F_2$ data agrees well with that determination based on the tagged
$D^\ast$ samples. The charm contribution to $F_2$ has been measured in an
extended kinematic ranges with improved precision. Experimental uncertainties
of these measurements are expected to be further reduced when all available
data will have been analysed. Both NC and CC cross-sections from $e^+p$ and 
$e^-p$ collisions at high $Q^2$ are measured. 
The electroweak effect has been observed for the first time at HERA 
in the $e^\pm p$ NC cross-sections. 
The different quark
contributions to the CC cross-sections have been experimentally established
providing potential for constraining the valence quark densities in the
future with more precise measurements. 

HERA is currently running with $e^+p$ collisions and will be upgraded in 2000 
with a factor of 5 increase in its peak luminosity and the possibility of
having polarised beams. These data provide an exciting opportunity both
in further precision tests of the Standard Model and in searching for new
physics beyond it.


\begin{thebibliography}{9}
\bibitem{h1newf2}H1 Collab., See e.g.\ M.~Klein, Proceedings of the 19th
International Symposium on Lepton and Photon Interactions at High-Energies,
Aug.9-14, 1999, Stanford, USA.
\bibitem{nmc}NMC Collab., M.~Arneodo el al., Nucl.\ Phys.\ B483 (1997) 3; 
             BCDMS Collab., A.~C.~Benvenuti et al., 
	     Phys. Lett.\ B223 (1989) 485; Phys.\ Lett.\ B237 (1990) 592.
\bibitem{dglap}G.~Altarelli and G.~Parisi, Nucl.\ Phys.\ B126 (1977) 298.
\bibitem{f2_zeus_94} ZEUS Collab., J.~Breitweg et al., Eur.\ Phys.\ J.\ C7
             (1999) 609.
\bibitem{ds_xs} H1 Collab., C.~Adloff et al., Nucl.\ Phys.\ B545 (1999) 21.
\bibitem{ds_pred} B.~W.~Harris and J.~Smith, Phys.\ Rev.\ D57 (1998) 2806.
\bibitem{ds_94} H1 Collab., C.~Adloff et al., Z.\ Phys.\ C72 (1996) 593; 
                ZEUS Collab., J.~Breitweg et al., Phys.\ Lett.\ B407 (1997)
                402.
\bibitem{nc_hiq2_e+-p} ZEUS Collab., J.~Breitweg et al., DESY 99-056;
                Contribution paper \#549 to HEP99;
                H1 Collab., C.~Adloff et al., DESY 99-107; Contribution
                paper \#157b to HEP99.
\bibitem{excess_hiq2} H1 Collab., C.~Adloff et al., Z.\ Phys.\ C74 (1997) 191;
                ZEUS Collab., J.~Breitweg et al., Z.\ Phys.\ C74 (1997) 207.
\bibitem{cc_hiq2_e+-p} H1 Collab., C.~Adloff et al., DESY 99-107; Contribution
                paper \#157b to HEP99; ZEUS Collab., J.~Breitweg et al., DESY
                99-059; Contribution paper \#558 to HEP99.
\bibitem{mw} PDG~98, C.~Caso et al., Eur.\ Phys.\ J.\ C3 (1998) 1.
\end{thebibliography}
\end{document}